\documentclass{ws-ijmpcs}
\usepackage{url}

\newcommand{\gevsq}{GeV$^2$}

\newcommand{\qsq}{$Q^2$}

\newcommand{\FD}{$ {F_{_{Dipole}}}$}

\newcommand{\GEp} {$ {G_{_{\rm E}} ^{\it p}}$}
\newcommand{\GMp}{$ {G_{_{\rm M}}^{\it p}}$}
\newcommand{\GEn} {$ {G_{_{\rm E}} ^{\it n}}$}
\newcommand{\GMn}{$ {G_{_{\rm M}}^{\it n}}$}

\newcommand{\Fone}{${ F_1}$}

\newcommand{\Fonep}{\mbox{${ {F_1^{\it p}}}$}}
\newcommand{\Ftwop}{\mbox{${ {F_2^{\it p}}}$}}

\newcommand{\Foneu}{\mbox{${ {F_1^{\it u}}}$}}

\newcommand{\Foned}{\mbox{${ {F_1^{\it d}}}$}}

\newcommand{\PR}{{ Phys. Rev. }}
\newcommand{\PRL}{{ Phys. Rev. Lett. }}

\newcommand{\etal}{{\em et al.}}

\begin{document}

\markboth{Bogdan Wojtsekhowski, Nucleon Form Factors }
{}
\catchline{}{}{}{}{}
\title{Nucleon form factors program with SBS at JLAB}

\author{Bogdan Wojtsekhowski}

\address{Thomas Jefferson National Accelerator Facility \\
Newport News, Virginia 23606 USA}

\maketitle

\begin{history}
\received{Day Month Year}
\revised{Day Month Year}
\end{history}

\begin{abstract}
The physics of the nucleon form factors is a fundamental part of the Jefferson Laboratory program.
We review the achievements of the 6-GeV era and the program with the 12-GeV beam
with the SBS spectrometer in Hall A, with a focus on the nucleon ground state properties.
\end{abstract}

\ccode{PACS numbers:14.20.Dh, 13.40.Gp}

\section{The high \qsq~nucleon form factor experiments}

The nucleon structure investigation using high energy electron scattering has been a successful 
field where many discoveries have been made since the 1956 observation of the proton form factor\cite{hof56}.
To a large extent, this success has been due to the dominance of the one-photon exchange mechanism of
electron scattering, which allows reliable interpretation of the experimental data\cite{ros50}.
By the early 90s, the form factor data sets for the proton and the neutron were found to be 
mainly proportional to the one form factor,  \FD$\,=\, (1+Q^2/0.71 [GeV^2])^{-2}$~for all four: 
magnetic and electric for the proton, and magnetic and electric for the neutron\cite{bos95}. 

The most decisive studies of the partonic structure of nucleon could be performed when the dominant 
part of the wave function is a 3-quark Fock state. 
This requires large momentum transfer, $Q^2 > 1$ GeV$^2$, when the contribution of the pion cloud is suppressed.
The SLAC experimental data\cite{arn86} on the proton Dirac form factor $F_1^p$ have been found 
to be in fair agreement with a scaling prediction\cite{lep79} based on perturbative QCD: \Fonep$\, 
\propto Q^{-4}$,  where \qsq~is the negative four-momentum transfer squared.

\begin{figure}[ht]
\unitlength 1cm
\begin{minipage}[th]{6.5cm}
   \centering
   \includegraphics[trim = 10mm 10mm 0mm 5mm, width=1.0\textwidth] {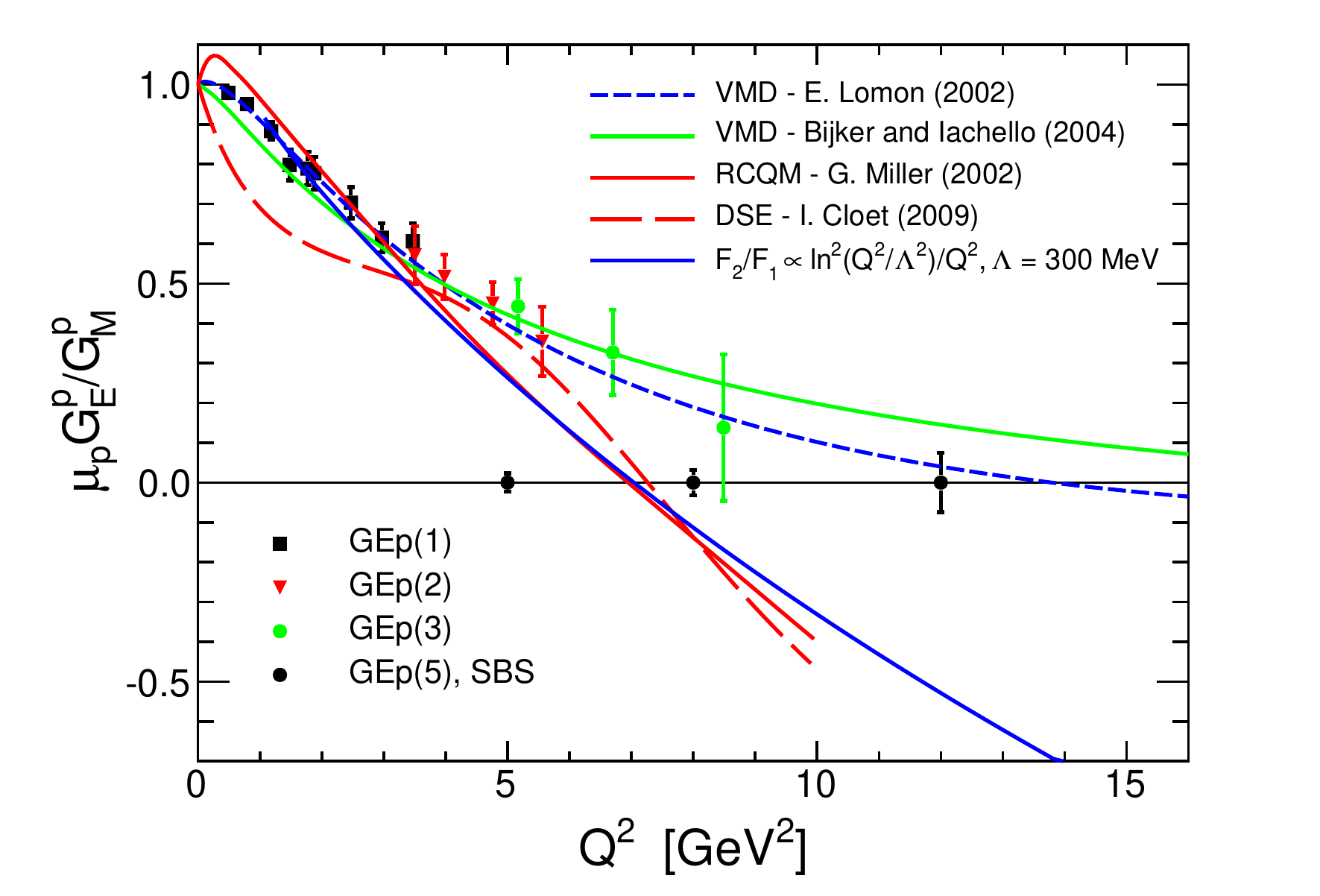} 
\end{minipage}
\hfill
\begin{minipage}[th]{5.7cm}
\includegraphics[trim = 35mm 15mm 35mm 35mm,width=1.0\textwidth] {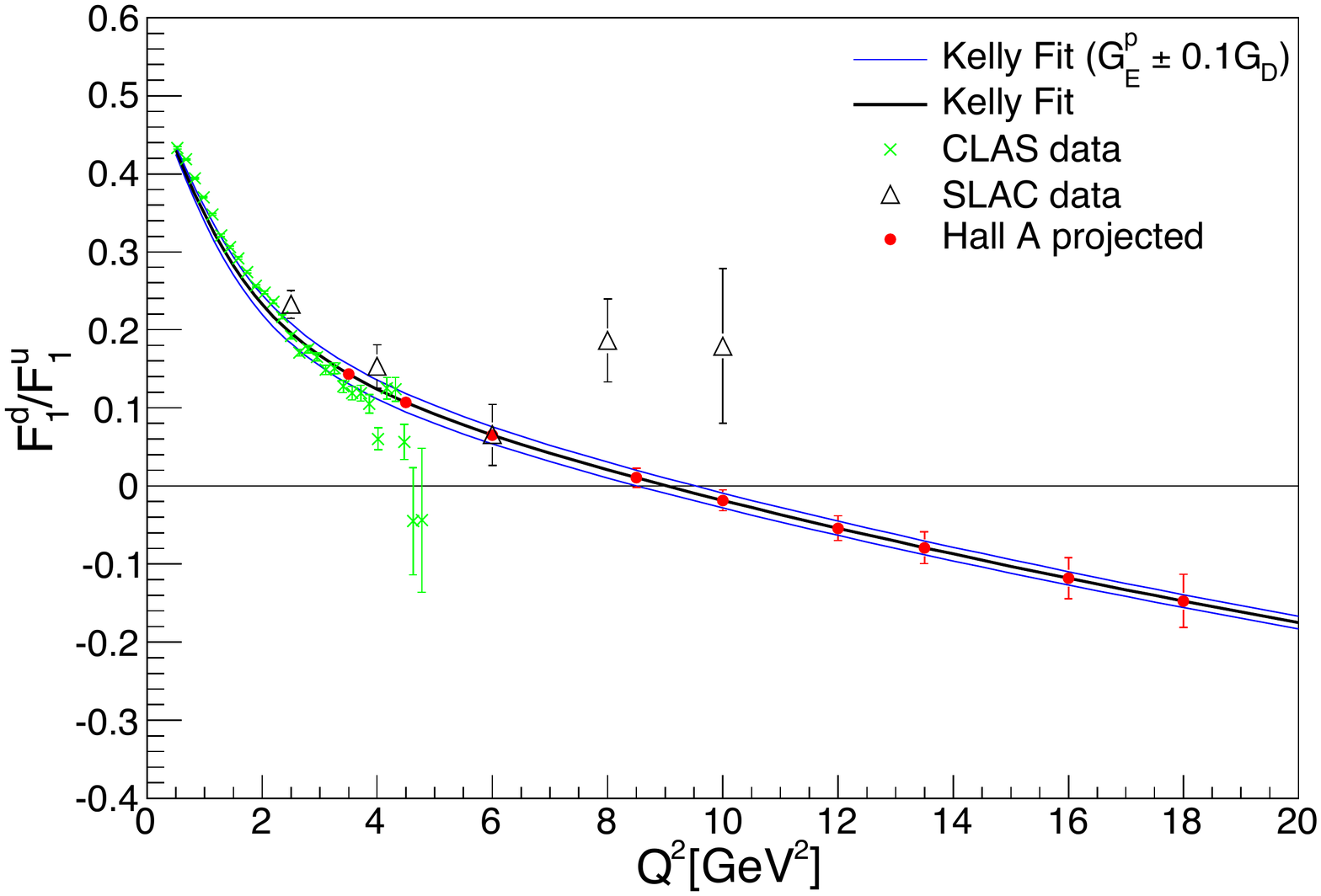}
\end{minipage}
\hfill
   \caption{ Left: Existing data and projected data accuracy for the ratio of the $\mu _p$\GEp/\GMp. 
Right: Ratio of the $\bf up$ and $\bf down$ quark contributions to the proton form factor \Fonep. } 
   \label{fig:GEp5}
\end{figure}

The experimental results\cite{jon00} from Jefferson Laboratory  (JLab) for the ratio of 
the proton Pauli form factor \Ftwop\ and the Dirac form factor \Fonep\ have been found to be 
in disagreement with the scaling law $F_2^p/F_1^p \propto 1/Q^2$ suggested in reference\cite{lep79}.  
A JLab high precision experiment made use of the double polarization method, which was 
first proposed in reference\cite{akh58}.
This method is less sensitive to the two-photon exchange contribution and, 
due to the interference nature of the double polarization asymmetry, has large sensitivity to the small electric form factor.
The data for $\mu_p$\GEp/\GMp~shown in Fig.~\ref{fig:GEp5}(left) present 
an amazing drop of \GEp, which also means that \Fonep~and \qsq$\times$\Ftwop~for 
the proton have different $Q^2$ dependencies.

The measurement of the proton to the neutron cross section ratio in the quasi-elastic knockout from the deuteron 
was used in JLab's precision measurement of the neutron magnetic form factor\cite{lac09}.
With the recent JLab experiment on the neutron electric form factor\cite{rio10}, the data on all four
nucleon form factors have become available in the \qsq~region of 3-quark dominance. 
Analysis of the flavor contributions to the nucleon form factors using the data was performed\cite{cat11}.
The flavor decomposition allowed us to make two new observations:
\begin{itemize}
\item The contributions of the {\bf up} quarks and {\bf down} quark to the magnetic and electric form 
factors of the proton all have different \qsq~ dependencies. 
\item The contribution of the {\bf down} quark to the \Fonep~form factor at \qsq=3.4~\gevsq~
is three times less than the contribution of the {\bf up} quarks (corrected for the number of quarks and 
their charge).
\end{itemize}

The second observation suggests that the probability of proton survival after the absorption of a massive virtual
photon is much higher when the photon interacts with an {\bf up} quark, which is doubly represented in the proton. 
This may be interpreted as an indication of an important role of the {\bf up-up} correlation.
At high \qsq~a correlation usually enhances the high momentum component and the interaction cross section.
The relatively weak {\bf down} quark contribution to the \Fonep~indicates a suppression of the {\bf up-down} correlation or a mutual cancellation of different types of {\bf up-down} correlations.  The QCD-based calculations of the nucleon form factors in the Dyson-Schwinger Equations approach\cite{rob07} revealed a key role of the diquark in high \qsq~electron-nucleon elastic scattering.

\section{Future experiments in Hall A with SBS}
\vskip -.15 in
Accurate measurement of the FFs at large \qsq~will be possible during the next few years at Jefferson Lab, where the 12-GeV energy upgrade is almost completed\cite{Jlab}.  
In 2007 we proposed a configuration of a large acceptance spectrometer at a small angle to 
the beam\cite{SBS}.  
This large luminosity moderate acceptance spectrometer, SBS, become a key component of the form factor
program in Hall A at JLab.
The program includes the measurements of three ratios: the proton electric form factor
to the proton magnetic form factor\cite{GEpE},  the neutron magnetic form factor 
to the proton magnetic form factor\cite{GMnE}, and the neutron electric form factor to the neutron magnetic form factor\cite{GEnE}. 
For absolute normalization of the form factor values, 
the precision measurement of the proton magnetic
form factor will also be performed in Hall A\cite{GMpE}. 
A summary of experimental parameters is shown in Table~\ref{tab:T1}.
\begin{table}[htb]
\tbl{Future measurements of the FFs in Hall A at JLab (approved experiments). 
Projected range of \qsq~and accuracy relatively the dipole FF at maximum value of \qsq.}
{\begin{tabular}{@{}cccl@{}} \toprule
Form factor & Reference & \qsq~range, \gevsq & $\Delta G/$\FD (stat/syst) at max \qsq \\ \colrule
\GEp   & \cite{GEpE} &    5-12       &   0.08 / 0.02   \\
\GMp   & \cite{GMpE} &  4.8-14.0  &   0.01 / 0.02  \\
\GEn   & \cite{GEnE}  &   1.5-10.2  &  0.09 / 0.03      \\
\GMn   & \cite{GMnE} &   3.5-13.5  &  0.06 / 0.03      \\  \botrule
\end{tabular} \label{tab:T1}}
\vskip -.25 in
\end{table}
\section{Flavor decomposition of the form factor \Fonep~and GPDs at very large \qsq}
At \qsq~above 10~\gevsq, measurement of the electric form factors, especially the \GEn, becomes difficult. 
However, due to the large value of $Q^2/4M_N^2$, the \Fone~could be obtained with a relatively
small uncertainty just from the value of the magnetic form factor:
\begin{equation}
F_1 \,=\, (G_{_E} + Q^2/4M_N^2 \times G_{_M})/(1+Q^2/4M_N^2)
\end{equation}
The flavor decomposition of \Fone~also could be accomplished accurately.
Fig.~\ref{fig:GEp5}(right) shows projected data points and a systematic error corridor
for assumed uncertainty in \GEp~of $\pm 0.1$.  
Here we used the Kelly fit form factors for illustration purposes.
We would like to note that the ratio \Foned/\Foneu~could potentially cross the zero line, which 
would require a significant change of GPDs parametrization because the currently used form does not allow negative values of GPDs, see e.g. the reference\cite{die13}.  

\section*{Acknowledgments}

The author would like to thank the organizing committee of the 
"International Workshop on e+e- collisions from Phi to Psi" for the invitation to present a talk.   This work was supported in part by the U. S. National Science Foundation, and by Department of Energy (DOE) contract number DE-AC05-06OR23177, under which the Jefferson Science Associates operates the Thomas Jefferson National Accelerator Facility.

\end{document}